\documentclass[showpacs,preprintnumbers,superscriptaddress]{revtex4}
\usepackage{CJK}
\usepackage{amsmath,amssymb,graphicx,bm}
\begin{document}

\title{Baryogenesis in quantum fluctuation modified gravity}

\author{Rong-Jia Yang \footnote{Corresponding author}}
\email{yangrongjia@tsinghua.org.cn}
\affiliation{College of Physics Science and Technology, Hebei University, Baoding 071002, China}
\affiliation{Hebei Key Lab of Optic-Electronic Information and Materials, Hebei University, Baoding 071002, China}
\affiliation{National-Local Joint Engineering Laboratory of New Energy Photoelectric Devices, Hebei University, Baoding 071002, China}
\affiliation{Key Laboratory of High-pricision Computation and Application of Quantum Field Theory of Hebei Province, Hebei University, Baoding 071002, China}

\author{Yong-Ben Shi}
\affiliation{College of Physics Science and Technology, Hebei University, Baoding 071002, China}

\begin{abstract}
We consider baryogenesis in quantum fluctuation modified gravity. We explore three forms (two are newly proposed here) of baryogenesis interaction and discuss the effect of these interaction terms on the baryon-to-entropy ratio during the radiation era of the expanding universe. We constrain the model parameters with the current observational data, implying that this modified gravity is capable to address the issue of baryon asymmetry in a successful manner.
\end{abstract}

\pacs{04.50.Kd, 98.80.-k; 04.60.-m; 04.90.+e}

\maketitle

\section{Introduction}
The problem of an excess of the matter over the anti-matter in the universe is one of the biggest mysteries of contemporary cosmology. This asymmetry is usually refer to the baryon asymmetry. The baryon number of the universe is precisely determined by observations, see for example, the latest result from Planck 2018 data \cite{Planck:2018vyg}
\begin{equation}
\eta_{\rm{B}}\equiv\frac{n_{\rm{B}}}{s}=(8.70 \pm 0.04)\times 10^{-11},
\end{equation}
where $n_{\rm B}$ and $s$ are the baryon number density and the entropy density of the present universe, respectively. Studies suggest that the asymmetry between matter and anti-matter might occur during the radiation eras \cite{Riotto:1999yt, Dine:2003ax, Cline:2006ts}. Sakharov suggested three conditions that are
necessary to generate a baryon asymmetry \cite{Sakharov:1967dj}: (i) non-conservation of baryon number; (ii) $\mathcal{C}$ and $\mathcal{CP}$ violation; (iii) deviation
from thermal equilibrium. Various baryogenesis scenarios, such as GUT Baryogenesis \cite{Kolb:1996jt}, Affleck-Dine
Baryogenesis \cite{Affleck:1984fy,Stewart:1996ai,Yamada:2015xyr,Akita:2017ecc}, electroweak
Baryogenesis \cite{Trodden:1998ym,Morrissey:2012db}, spontaneous Baryogenesis \cite{Takahashi:2003db,DeSimone:2016ofp}, baryogenesis through evaporation of primordial black
holes \cite{Dolgov:1980gk} or via spontaneous Lorentz violation \cite{Carroll:2005dj}, and etc \cite{Alexander:2004us, Mohanty:2005ud, Abbott:1982hn, Li:2004hh, Hamada:2007vu, Shiromizu:2004cb, Cohen:1987vi, Fukugita:1986hr, Feng:2022gdz, Li:2002wd, Goodarzi:2023ltp, Cado:2023gan, RufranoAliberti:2023ywq,Kawai:2017kqt,MohseniSadjadi:2007qk,Oikonomou:2015qfh}, had been suggested to explain how there are more matter than antimatter in the universe.

Recently, a model had been suggested to generate the baryon number asymmetry by violation of $\mathcal{CPT}$, while the
thermal equilibrium is maintained \cite{Cohen:1987vi}. In \cite{Davoudiasl:2004gf}, a dynamical violation of $\mathcal{CPT}$ (and also $\mathcal{CP}$) may give
rise to the baryon asymmetry in thermal equilibrium during the expansion of the
universe. The interaction responsible for $\mathcal{CP}$ violation is given by a coupling between the derivative of the Ricci
scalar and the baryon current. This baryogenesis, called Gravitational baryogenesis, had been generalised to $f(R)$ theories \cite{Lambiase:2006dq}, to Gauss-Bonnet gravity \cite{Odintsov:2016hgc}, to $f(R,T)$ theories \cite{Baffou:2018hpe,Nozari:2018ift,Sahoo:2019pat}, to $f(T)$ theory \cite{Oikonomou:2016jjh}, to $f(R, L_{\rm{m}})$ theory \cite{Jaybhaye:2023lgr}, to loop quantum gravity \cite{Odintsov:2016apy}, and so on.

To explain the accelerated expansion of the universe, modified gravity is one popular attempt, such as $f(R)$ theories, Gauss-Bonnet gravity, and $f(R, T)$ theory. Recently, a type of modified gravity was proposed from Heisenberg's nonperturbative quantization \citep{Dzhunushaliev:2013nea,Dzhunushaliev:2015mva,Dzhunushaliev:2015hoa}: if the metric can be decomposed as the sum of classical and quantum fluctuating parts, then the corresponding Einstein quantum gravity generates at the classical level modified gravity models with a non-minimal coupling between geometry and matter \cite{Yang:2015jla}. This idea has been realized in some specific models  \cite{Yang:2015jla,Liu:2016qfx,Bernardo:2021ynf,Chen:2021oal,Lima:2023qdv}. Some of these modified gravity had been applied to black hole physics \cite{Yang:2020lxv} and to Friedmann-Lemaitre-Robertson-Walker geometries \cite{Yang:2015jla,Liu:2016qfx,Bernardo:2021ynf,Haghani:2021iqe,Chen:2021oal,Kawai:2017kqt}. Since the quantum fluctuations of metric will lead to modified gravitational field equations, it may cause the symmetry breaking between matter and anti-matter. In this paper, we will investigate this issue in the framework of quantum fluctuation modified gravity (QFMG) proposed in \cite{Yang:2015jla}. The results show that the QFMG may provide a possible solution to the problem of baryon asymmetry.

The paper is organized as follows. In the next Section, we will briefly review the QFMG proposed in \cite{Yang:2015jla}. In Sec. III, we will discuss the baryon asymmetry in this modified gravity. Finally, we will briefly summarize and discuss our results in section IV. For numerical calculations
and graphs, the physical constants take the following values: $m_{\rm p}\sim 1.22\times 10^{19}$ GeV, $g_{\rm{b}}\simeq \mathcal{O}(1)$, $g_*\simeq 106$, $\eta_{\rm{B}}=8.70 \times 10^{-11}$, and $G_0\sim 1/m^2_{\rm p}$.

\section{Quantum fluctuation modified gravity}
If the metric operator can be decomposed into a sum of a classical metric $g_{\mu\nu}$ and a fluctuating part $\widehat{\delta g}_{\mu\nu}$ \citep{Dzhunushaliev:2013nea}: $\hat{g}_{\mu\nu}=g_{\mu\nu}+\widehat{\delta g}_{\mu\nu}$, where the classical metric can be seen as the background field, so the quantum state will not be vacuum state, in other words the expectation value of the quantum part of the metric operator should not vanish: $\langle\delta \hat{g}_{\mu\nu}\rangle\neq 0$. Ignoring high order fluctuations, the quantum Einstein-Hilbert Lagrangian $L_{\hat{g}}=\frac{1}{2k^2}\sqrt{-\hat{g}} \hat{R}$ can be expanded as
\begin{eqnarray}
\label{lag0}
L_{\hat{g}}=L_{\hat{g}}(g+\delta\hat{g})\approx L_{g}(g)+\frac{\delta L_{g}}{\delta g^{\mu\nu}}\widehat{\delta g}^{\mu\nu}.
\end{eqnarray}
Hereafter $k^2=8\pi G$ and we take $c=\hbar=k_B=1$ with $k_{\rm{B}}$ the Boltzmann constant. Since $\left\langle \frac{\delta L_{g}}{\delta g^{\mu\nu}}\widehat{\delta g}^{\mu\nu}\right\rangle=\frac{\delta L_{g}}{\delta g^{\mu\nu}}\langle \widehat{\delta g}^{\mu\nu}\rangle=\sqrt{-g}G_{\mu\nu}\langle \widehat{\delta g}^{\mu\nu}\rangle$ \cite{Dzhunushaliev:2013nea}, the expectation value of Lagrangian (\ref{lag0}) has the form
\begin{eqnarray}
\label{lag00}
\langle L_{\hat{g}}\rangle\approx \frac{1}{2k^2}\sqrt{-g} \big[R+G_{\mu\nu}\langle\delta \hat{g}^{\mu\nu}\rangle \big],
\end{eqnarray}
Similarly, the quantum Lagrange density $L^{\hat{g}}_{\rm m}$ can be expanded as follows
\begin{eqnarray}
\label{lagm}
L^{\hat{g}}_{\rm m}(g+\delta\hat{g})\approx \sqrt{-g}L_{\rm m}(g)+\frac{\delta \sqrt{-g}L_{\rm m}}{\delta g^{\mu\nu}}\widehat{\delta g}^{\mu\nu}.
\end{eqnarray}
With the aforementioned assumptions, the expectation value of the quantum Lagrange density (\ref{lagm}) is given by \citep{Dzhunushaliev:2013nea}
\begin{eqnarray}
\label{lagm0}
\langle L^{\hat{g}}_{\rm m}(g+\delta\hat{g}) \rangle \approx \sqrt{-g}\big[L_{\rm m}-\frac{1}{2}T_{\mu\nu}\langle\delta \hat{g}^{\mu\nu}\rangle\big],
\end{eqnarray}
where $T_{\mu\nu}=-2\delta(\sqrt{-g}L_{\rm m})/(\sqrt{-g}\delta g^{\mu\nu})$ is the energy-momentum tensor, Then the modified Lagrangian density can be written as
\begin{eqnarray}
\label{lag}
L=\frac{1}{2k^2}\sqrt{-g} \big[R+G_{\mu\nu}\langle\delta \hat{g}^{\mu\nu}\rangle \big]+\sqrt{-g}\big[L_{\rm m}-\frac{1}{2}T_{\mu\nu}\langle\delta \hat{g}^{\mu\nu}\rangle\big].
\end{eqnarray}
The specific forms of modified Lagrangian density \eqref{lag} depend on choices of quantum fluctuations $\delta \hat{g}^{\mu\nu}$. In \cite{Wetterich:2016vxu}, $\delta \hat{g}^{\mu\nu}$ is split into a "physical metric" and a "gauge part". By a suitable gauge transformation one can eliminate the gauge part. In general, $\delta \hat{g}^{\mu\nu}$ may depend on the background field $g_{\mu\nu}$ or the matter field $T_{\mu\nu}$, or it may be completely independent of them. In \cite{Yang:2015jla}, a simple case $\langle\delta \hat{g}^{\mu\nu}\rangle=\alpha g^{\mu\nu}$ with $\alpha$ a constant has been considered, which fulfills two conditions, $\nabla_\alpha\langle\delta \hat{g}^{\mu\nu}\rangle=0$ and $\langle\delta \hat{g}^{\mu\nu}\rangle=\langle\delta \hat{g}^{\nu\mu}\rangle$, suggested in \cite{Wetterich:2016vxu}, then the Lagrangian density (\ref{lag}) becomes as
\begin{eqnarray}
\label{act}
L=L_{\rm mg}+L_{\rm mm}=\frac{1}{2k^2}\sqrt{-g}(1-\alpha)R+\sqrt{-g}[L_{\rm m}-\frac{1}{2}\alpha T],
\end{eqnarray}
where $T=g_{\mu\nu}T^{\mu\nu}$ is the trace of the energy-momentum tensor. Here we assume that the quantum fluctuation is less than the classical part of the metric operator: $|\alpha|<1$. Varying the Lagrangian density (\ref{lag}) with respect to $g^{\mu\nu}$ and assuming $\delta g_{\mu\nu}=0$ on the boundary, we obtain the gravitational field equations \cite{Yang:2015jla}
\begin{eqnarray}
\label{mot}
R_{\mu\nu}-\frac{1}{2}g_{\mu\nu}R=\frac{2k^2}{1-\alpha}\left[\frac{1}{2}(1+\alpha)T_{\mu\nu}-\frac{1}{4}\alpha g_{\mu\nu}T+\frac{1}{2}\alpha\theta_{\mu\nu}\right],
\end{eqnarray}
where $\theta_{\mu\nu}=g^{\alpha\beta}\delta T_{\alpha\beta}/\delta g^{\mu\nu}$ and $\theta=g_{\mu\nu}\theta^{\mu\nu}$, for detail see \cite{Yang:2015jla}. For convenience, we call $\theta_{\mu\nu}$ as energy-momentum induced tensor. We observe that the gravitational constant $G$ and the energy-momentum tensor are modified due to the quantum fluctuation of the metric. Since $R=-k^2[T+\alpha\theta/(1-\alpha)]$, the gravitational field equations (\ref{mot}) can be rewritten as \cite{Yang:2015jla}
\begin{eqnarray}
\label{mot1}
R_{\mu\nu}=\frac{2k^2}{1-\alpha}\left[\frac{1}{2}(1+\alpha)T_{\mu\nu}-\frac{1}{4} g_{\mu\nu}T+\frac{1}{2}\alpha\theta_{\mu\nu}-\frac{1}{4}\alpha g_{\mu\nu}\theta\right].
\end{eqnarray}
Thinking of $\nabla_{\nu}G^{\mu\nu}=0$, we derive the equation for the divergence of the stress-energy tensor $T_{\mu\nu}$ \cite{Yang:2015jla}
\begin{eqnarray}
\label{conser}
\nabla^{\nu}T_{\mu\nu}=\frac{1}{1+\alpha}\left[\frac{1}{2}\alpha \nabla_{\mu}T-\alpha\nabla^{\nu}\theta_{\mu\nu}\right].
\end{eqnarray}
In other words, if taking into account of quantum fluctuations, the stress-energy tensor are not conserved quantities any more. Discussions about the nonconservation of stress-energy tensor can be found in \citep{Bertolami:2008ab,Harko:2008qz,Bisabr:2012tg,Minazzoli:2013bva}. As pointed out in \cite{Liu:2016qfx,Yang:2020lxv}, the Lagrangian density (\ref{lag}) suggests an possible equivalent microscopic quantum description of the matter creation processes in $f(R, T)$ or $f(R, L_{\rm{m}})$ gravity. Such a description may shed some lights on the physical mechanisms leading to particle generation via gravity and matter geometry coupling.

\section{Baryogenesis in quantum fluctuation modified gravity}
Consider a homogeneous and isotropic Friedmann-Lema\^{i}tre-Robertson-Walker (FLRW) universe
\begin{align}
ds^2=dt^2-a^2(t)\left[\frac{dr^2}{1+Kr^2}+r^2(d\theta^2+\sin^2\theta d\phi^2)\right],
\end{align}
where $a$ is the scalar factor. The spatial curvature constant $K=+1$, 0, and $-1$ correspond to a closed, flat, and open universe, respectively. And suggest a perfect fluid specified by the energy-momentum tensor
\begin{eqnarray}
T_{\mu\nu}=(\rho+p)u_{\mu}u_{\nu}-g_{\mu\nu}p.
\end{eqnarray}
where the fluid 4-velocity fulfil the condition $u^{\mu}u_{\mu}=1$. Taking the Lagrangian for matter as $L_{\rm m}=-p$, we have from the variation of the energy-momentum tensor of a perfect fluid
\begin{eqnarray}
\label{emit}
\theta_{\mu\nu}=-2T_{\mu\nu}-g_{\mu\nu}p.
\end{eqnarray}
With $T=\rho-3p$ and $\theta=-2T-4p=-2(\rho-p)$, the $\mu\nu=00$ component of Eq. (\ref{mot1}) is \cite{Yang:2015jla}
\begin{eqnarray}
\label{motf}
\frac{\ddot{a}}{a}=-\frac{1}{3}\frac{k^2}{2(1-\alpha)}\big[\rho+(3-4\alpha)p\big],
\end{eqnarray}
and the equations $\mu\nu=00$ and $\mu\nu=11$ together yield \cite{Yang:2015jla}
\begin{eqnarray}
\label{motf1}
H^2+\frac{K}{a^2}=\frac{1}{3}\frac{k^2}{2(1-\alpha)}\left[(2-3\alpha)\rho+\alpha p\right].
\end{eqnarray}
For a spatial flat FLRW spacetime, Eqs. (\ref{motf1}) and (\ref{motf}) become \cite{Yang:2015jla}
\begin{eqnarray}
\label{motfd1}
H^2&=&\frac{k^2}{3}\left[\frac{2-(3-w)\alpha}{2-2\alpha} \right]\rho,\\
\label{motfd2}
\dot{H}+H^2&=&\frac{\ddot{a}}{a}=-\frac{k^2}{6}\left[\frac{1+(3-4\alpha)w}{1-\alpha}\right]\rho,
\end{eqnarray}
with $w=p/\rho$. Contracting these two equations, we have
\begin{equation}
\dot{H}+ \frac{3(1-\alpha)\left(1+\omega\right)}{2+ \alpha\left(\omega-3\right)} H^2=0.
\end{equation}
Solving this equation, we get
\begin{equation}
\label{Hsol}
H=\frac{\lambda}{t},
\end{equation}
where
\begin{equation}
\lambda=\frac{2 -\left(3-w\right) \alpha}{3\left(1-\alpha \right)\left(1+w\right)}.
\end{equation}
Substituting Eq. (\ref{Hsol}) into Eq. (\ref{motfd1}), yields
\begin{equation}
\label{rhos}
\rho=\rho_0 t^{-2},
\end{equation}
where
\begin{equation}
\rho _{0}=\frac{2[2-(3-\omega)\alpha]}{3k^2(1-\alpha)(1+\omega)^{2}}.
\end{equation}
Assuming that he universe evolves slowly from an equilibrium state to an equilibrium state, the energy density links to the
temperature $\mathcal{T}$ as
\begin{equation}
\label{rot}
\rho=\frac{\pi^{2}}{30} g_{*} \mathcal{T}^{4},
\end{equation}
where $g_{*}$ denotes the number of the degrees of freedom of the effectively massless particles. Using this equation and the energy density \eqref{rhos}, we find the decoupling time as a function of decoupling temperature $\mathcal{T}_{\rm{D}}$ as
\begin{equation}
\label{tt}
t_{\rm{D}}=\left(\frac{30 \rho_{0}}{\pi^{2} g_{*}}\right)^{\frac{1}{2}} \mathcal{T}_{\rm{D}}^{-2}.
\end{equation}
As shown in \cite{Yang:2015jla}, the inflationary era predictions is well motivated and is viable to stand as a viable theoretical framework of inflation in QFMG.
With Eq. \eqref{tt} and a certain baryogenesis interaction term, we can discuss whether this mechanism generates baryon asymmetry during the expansion of the universe.

\subsection{Coupling of the baryon current with Ricci scalar}
\begin{figure}
\centering
\includegraphics[height=8cm,width=10cm]{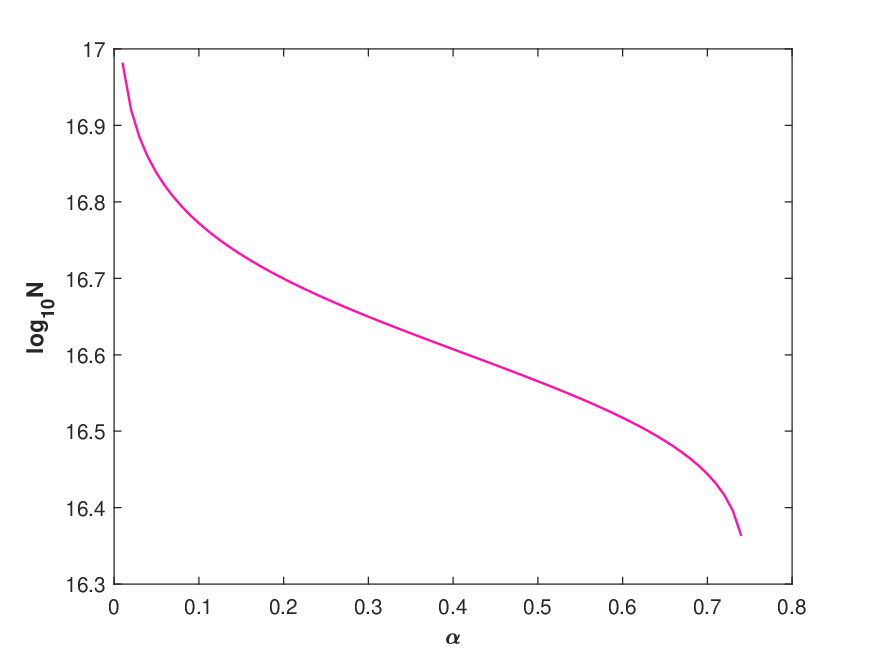}
\caption{The re-scaled decoupling temperature $N$ as a function of the parameter $\alpha$ in the coupling model $J_\mu \partial^\mu R$.}
\label{ma1}
\end{figure}

\begin{figure}
\centering
\includegraphics[height=8cm,width=10cm]{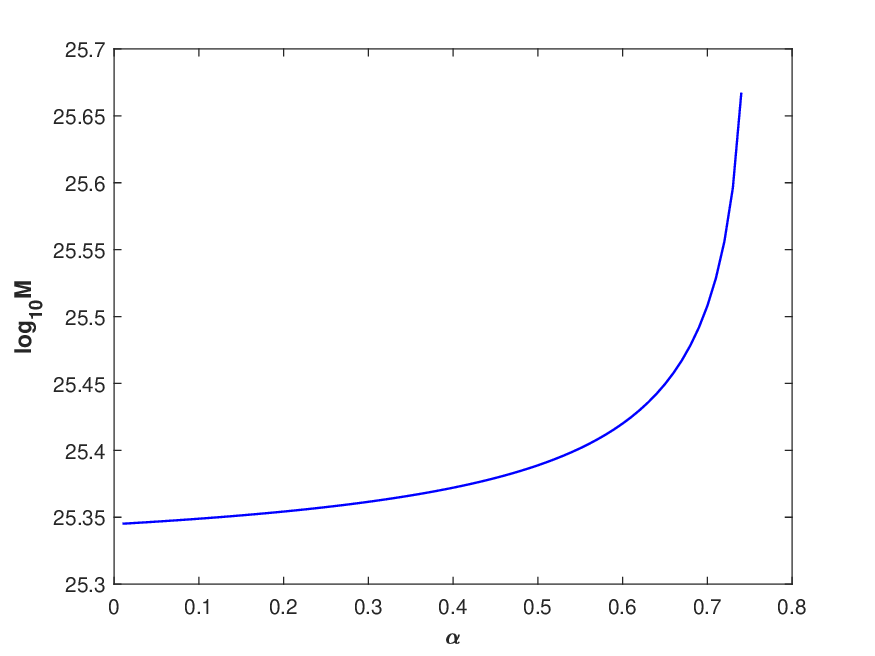}
\caption{The cutoff scale $M$ as a function of the parameter $\alpha$ in the coupling model $J_\mu \partial^\mu R$.}
\label{Ta1}
\end{figure}

We first consider the possibility that the baryon current couplings with Ricci scalar. The interaction between the derivative of the Ricci scalar
$R$ and the baryon-number current $J_\mu$ breaks $\mathcal{CPT}$ dynamically in the expansion of the universe. The $\mathcal{CP}$-violating interaction term is given by
\begin{equation}
\label{Rc}
\frac{1}{M_{*}^{2}} \int d^4 x \sqrt{-g} J^{\mu} \partial_{\mu} R,
\end{equation}
where $M_*$ is the cutoff scale of underlying effective theory. After the temperature of the universe drops
below the critical temperature $\mathcal{T}_{\rm{D}}$, the baryon-to-entropy ratio in the expansion of universe becomes \cite{Davoudiasl:2004gf}
\begin{equation}
 \label{nb1}
\frac{n_{\rm{B}}}{s} \simeq -\frac{15 g_{\rm{b}}}{4 \pi^{2} g_{*}} \frac{\dot{R}}{M_{*}^{2}\mathcal{T}_{\rm{D}}}.
\end{equation}
For a flat FLRW metric, the Ricci scalar is given by
\begin{equation}
\label{rt}
R=-6\left(\dot{H}+2 H^{2}\right)=\frac{6\lambda(2\lambda-1)}{t^{2}},
\end{equation}
So the time derivative of the Ricci scalar is
\begin{equation}
\label{rdot}
\dot{R}=\frac{12\lambda(2\lambda-1)}{t^{3}}=\frac{12\lambda(2\lambda-1)}{t^{3}}.
\end{equation}
Inserting Eqs. \eqref{tt} and (\ref{rdot}) into Eq. (\ref{nb1}), we obtain the baryon asymmetry factor as
\begin{eqnarray}
\label{nbs1}
 \frac{n_{\rm{B}}}{s}\simeq -\left(\frac{648}{5}\right)^{\frac{1}{2}}\frac{ g_{\rm b}g^{1/2}_{*}\pi^{5/2}\lambda(2\lambda-1)}{M_{*}^{2}G^{-3/2}}\left[\frac{(1-\alpha)(1+w)^2}{2-(3-w)\alpha}\right]^{\frac{3}{2}}\mathcal{T}_{D}^{5}.
\end{eqnarray}
If $\lambda\neq 1/2$ (i.e. $\alpha\neq0$), this equation shows that the resulting baryon-to-entropy ratio is non-zero even in radiation dominated era ($w= 1/3$). Thinking of the conditions: $|\alpha|<1$ and $\eta_{\rm{B}}>0$, we have $0<\alpha<3/4$ from Eq. (\ref{nbs1}) for $w= 1/3$. Assuming that $g_{\rm{b}}\simeq \mathcal{O}(1)$ and $g_*\simeq 106$, there are still three parameters needed to be determined: $\alpha$, $M_{*}$, and $\mathcal{T}_{\rm{D}}$. Letting $M_{*}=xm_{\rm{p}}$ and $N=\mathcal{T}_{\rm{D}}/x^2$, we plot in figure \ref{ma1} the re-scaled temperature $N$ as a function of the parameter $\alpha$. We observe that the re-scale temperature $N$ decreases as the parameter $\alpha$ increasing.

Letting $\mathcal{T}_{\rm{D}}=ym_{\rm{p}}$ and $M=M_{*}/y^{5/2}$, we plot the cutoff scale $M$ as a function of the parameter $\alpha$ in figure \ref{Ta1}. We observe that the cutoff scale $M$ increases as the parameter $\alpha$ increasing. 

If $T_{\rm{D}}=M_{\rm{I}}$ where $M_{\rm{I}}\sim 2\times10^{16}$ GeV is the upper limit on the tensor mode fluctuation constraints in inflationary scale, a possible choice of the cutoff scale $M_{*}$ is $M_{*}=m_{\rm p}/\sqrt{8\pi}$, where $m_{\rm p}$ is the Planck mass \cite{Davoudiasl:2004gf}, we get $\alpha\simeq0.46$.

\subsection{Coupling of the baryon current with energy-momentum induced tensor}
\begin{figure}
\centering
\includegraphics[height=8cm,width=10cm]{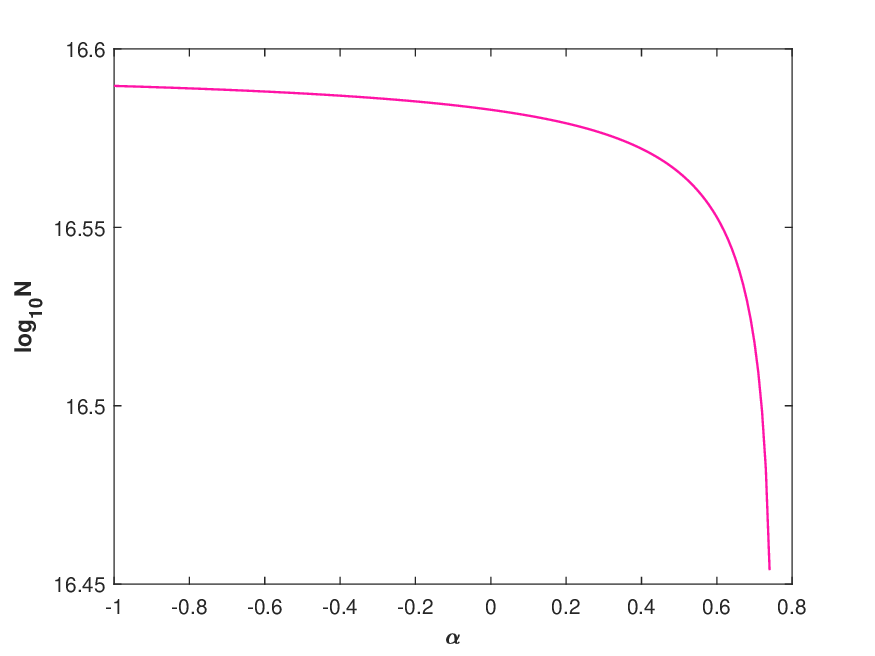}
\caption{The re-scaled decoupling temperature $N$ as a function of the parameter $\alpha$ in the coupling model $-J_\mu \partial^\mu \theta$.}
\label{ma21}
\end{figure}

\begin{figure}
\centering
\includegraphics[height=8cm,width=10cm]{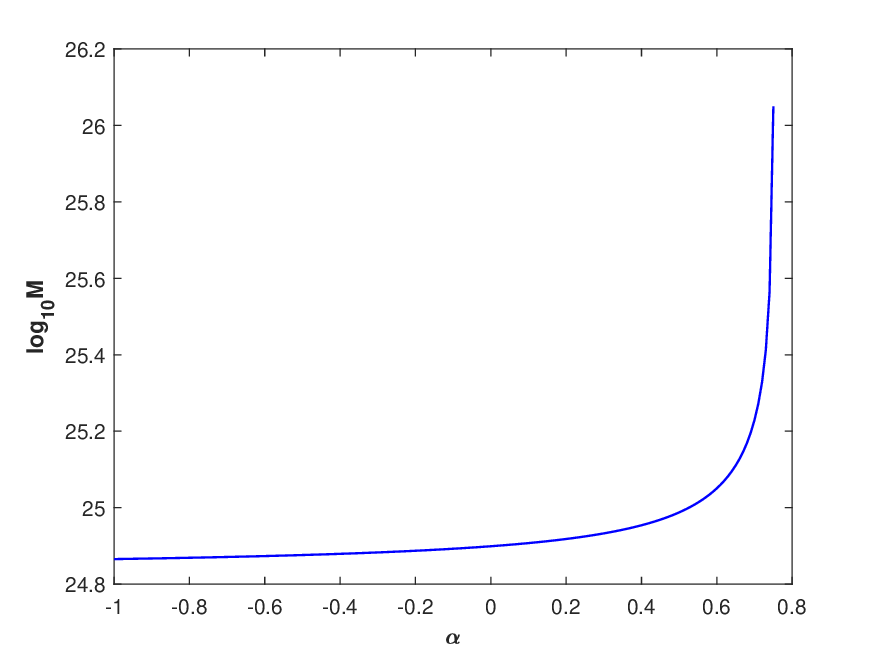}
\caption{The cutoff scale $M$ as a function of the parameter $\alpha$ in the coupling model $-J_\mu \partial^\mu \theta$.}
\label{Ta21}
\end{figure}

In some literatures, the baryogenesis interaction term, $J_{\mu}\partial^{\mu}T$, was considered. However, since $T=\rho-3p=(1-3w)\rho$, this interaction term will not contribute to the baryon asymmetry in the radiation dominated era. Here we consider a new baryogenesis interaction term which is given by
\begin{equation}
-\frac{k^2}{M_{*}^{2}} \int d^{4} x \sqrt{-g} J^{\mu} \partial_{\mu}\theta.
\end{equation}
For this case, the baryon-to-entropy ratio can be defined as
\begin{equation}
 \label{nb}
\frac{n_{\rm{B}}}{s} \simeq \frac{15 g_{\rm{b}}k^2}{4 \pi^{2} g_{*}} \frac{\dot{\theta}}{M_{*}^{2}\mathcal{T}_{\rm{D}}}.
\end{equation}
From Eq. (\ref{emit}), and combining with Eq. \eqref{rhos}, one can easily get
\begin{equation}
\theta=-2(\rho-p)=-2(1-w)\rho=-2(1-w)\rho_0t^{-2}.
\end{equation}
Combining this equation with Eq. \eqref{tt}, we derive
\begin{eqnarray}
\label{nbs2}
\frac{n_{\rm{B}}}{s}
\simeq\left(\frac{32}{5}\right)^{\frac{1}{2}}\frac{g_{\rm b}g^{1/2}_{*}\pi^{5/2}(1-w)}{M_{*}^{2}G^{-3/2}}
\left[\frac{(1-\alpha)(1+w)^2}{2-(3-w)\alpha}\right]^{\frac{1}{2}}\mathcal{T}_{D}^{5}.
\end{eqnarray}
For $w=1/3$, the conditions $|\alpha|<1$ and $\eta_{\rm{B}}>0$ imply that $-1<\alpha<3/4$ and $\alpha\neq 0$. As above, we plot in figure \ref{ma21} the re-scale temperature $N$ as a function of the parameter $\alpha$. We find that the re-scale temperature $N$ decreases as the parameter $\alpha$ increasing. The cutoff scale $M$ as a function of the parameter $\alpha$ is shown in figure \ref{Ta21}. We observe that the cutoff scale $M$ increases as the parameter $\alpha$ increasing. If taking $T_{\rm{D}}=M_{\rm{I}}\sim 2\times10^{16}$ GeV and $M_{*}=m_{\rm p}/\sqrt{8\pi}$, we have $\alpha\simeq0.12$.

\subsection{A generalized baryogenesis interaction term}
\begin{figure}
\centering
\includegraphics[height=8cm,width=10cm]{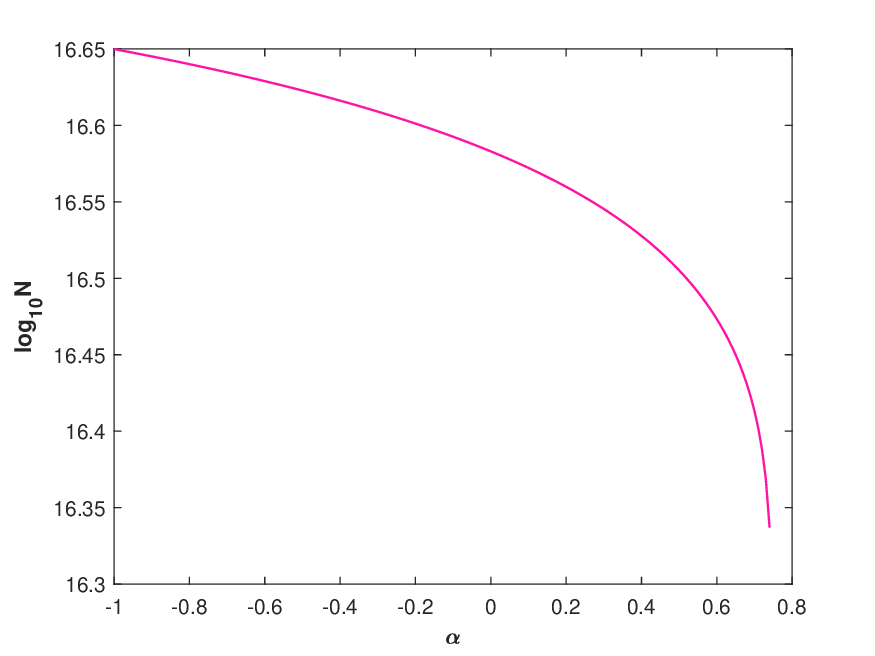}
\caption{The re-scaled decoupling temperature $N$ as a function of the parameter $\alpha$ in the coupling model $J_\mu (\partial^\mu R-k^2\partial^\mu \theta)$.}
\label{ma3}
\end{figure}
\begin{figure}
\centering
\includegraphics[height=8cm,width=10cm]{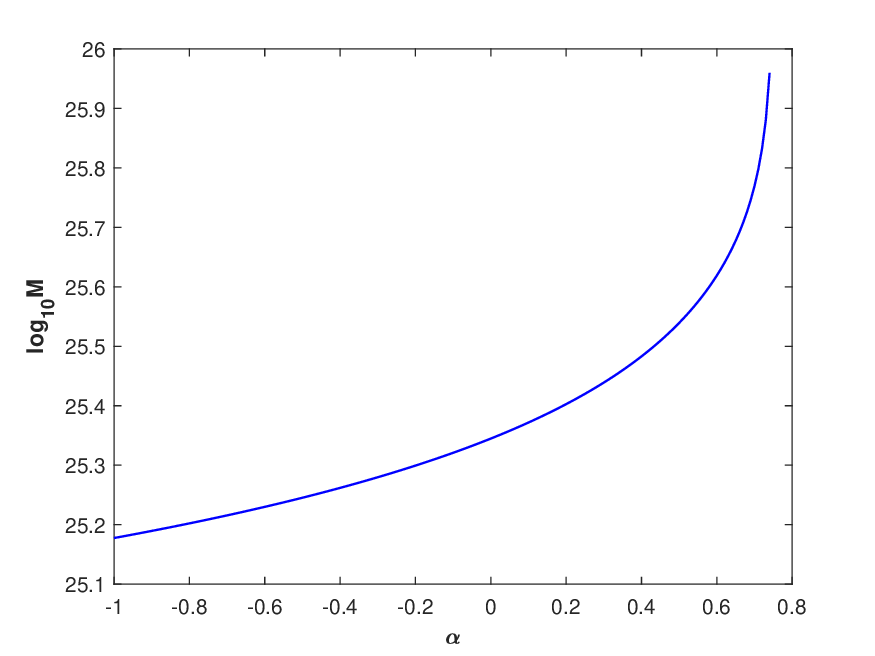}
\caption{The cutoff scale $M$ as a function of the parameter $\alpha$ in the coupling model $J_\mu (\partial^\mu R-k^2\partial^\mu \theta)$.}
\label{Ta3}
\end{figure}

Those two baryon generation mechanisms discussed above inspire we to consider a more general baryogenesis interaction term
\begin{equation}
\frac{1}{M^{2}_{*}} \int d^4 x \sqrt{-g} J^{\mu} \partial_{\mu} F(R, \theta) \ , \\ \label{28}
\end{equation}
where $F(R, \theta)$ is any function of $R$ and $\theta$. The baryon to entropy ratio can be calculated as
\begin{equation}
\frac{n_{\rm{B}}}{s}\simeq-\frac{15 g_{b}}{4 \pi^{2} g_{*}} \frac{\dot{R} F_{R}+\dot{\theta} F_{\theta}}{M_{*}^{2} \mathcal{T}_{D}} \ , \\ \label{29}
\end{equation}
where $F_{R}\equiv dF/dR$ and $F_\theta\equiv dF/d\theta$. Using Eqs. (\ref{nbs1}) and (\ref{nbs2}), we obtain
\begin{equation}
\label{nbs3}
\frac{n_{\rm{B}}}{s} \simeq \left\{-\left(\frac{648}{5}\right)^{\frac{1}{2}}\lambda(2\lambda-1)\left[\frac{(1-\alpha)(1+w)^2}{2-(3-w)\alpha}\right]^{\frac{3}{2}}F_{R}
+\left(\frac{32}{5}\right)^{\frac{1}{2}}(1-w)
\left[\frac{(1-\alpha)(1+w)^2}{2-(3-w)\alpha}\right]^{\frac{1}{2}} F_{\theta}\right\}\frac{g_{\rm b}g^{\frac{1}{2}}_{*}\pi^{\frac{5}{2}}\mathcal{T}_{D}^{5}}{M_{*}^{2}G^{-3/2}}.
\end{equation}
Since $F(R, \theta)$ is unknown, here we consider a simple case: $F(R, \theta)=R-k^2\theta$. Then from \eqref{nbs3}, we have
\begin{equation}
\label{nbs4}
\frac{n_{\rm{B}}}{s} \simeq \left\{-\left(\frac{648}{5}\right)^{\frac{1}{2}}\lambda(2\lambda-1)\left[\frac{(1-\alpha)(1+w)^2}{2-(3-w)\alpha}\right]^{\frac{3}{2}}
+\left(\frac{32}{5}\right)^{\frac{1}{2}}(1-w)
\left[\frac{(1-\alpha)(1+w)^2}{2-(3-w)\alpha}\right]^{\frac{1}{2}}\right\}\frac{g_{\rm b}g^{\frac{1}{2}}_{*}\pi^{\frac{5}{2}}\mathcal{T}_{D}^{5}}{M_{*}^{2}G^{-3/2}}.
\end{equation}
For $w=1/3$, the conditions $|\alpha|<1$ and $\eta_{\rm{B}}>0$ imply that $-1<\alpha<3/4$. In figure \ref{ma3}, we plot re-scale temperature $N$ as a function of the parameter $\alpha$. We observe that re-scale temperature $N$ decreases as the parameter $\alpha$ increasing. We plot in figure \ref{Ta3} the cutoff scale $M$ as a function of the parameter $\alpha$. We find that the parameter $M$ increases as the parameter $\alpha$ increasing. Taking $T_{\rm{D}}=M_{\rm{I}}\sim 2\times10^{16}$ GeV and $M_{*}$ is $M_{*}=m_{\rm p}/\sqrt{8\pi}$, we get $\alpha\simeq0.02$.

\section{Conclusions and discussions}
We have discussed baryogenesis scenario in quantum fluctuation modified gravity. We have explored that how this modified
gravity is capable to explain the observed asymmetry between matter and antimatter. We have considered three forms (two are newly proposed here) of baryogenesis interaction and investigate the effects of these interaction terms on the baryon-to-entropy ratio during the radiation era of the expanding universe. The parameter characterized the quantum fluctuation was constrained as: $-1<\alpha<3/4$, or $0<\alpha<3/4$. So even for $\mid\alpha\mid\ll 1$, these models can still generate baryon asymmetry. We have plotted the re-scaled decoupling temperature or the cutoff scale as a function of the model parameter and have presented some values of the parameter $\alpha$ in these coupling models when other parameters taking some special values. We have shown that the current observational value of the baryon-to-entropy ratio can be obtained for large set of parameters of the models as well as the decoupling temperature, and the characteristic cutoff scale.

\begin{acknowledgments}
This study is supported in part by National Natural Science Foundation of China (Grant No. 12333008) and Hebei Provincial Natural Science Foundation of China (Grant No. A2021201034).
\end{acknowledgments}
\bibliographystyle{elsarticle-num}
\bibliography{refq1}

\end{document}